\documentclass[prd,reprint,nofootinbib,amsmath,amssymb,aps,floatfix]{revtex4-1}
\usepackage{amssymb,amsmath,graphicx,color}
\usepackage[linktoc=page]{hyperref}
\usepackage{subfigure}
\usepackage{amsfonts}

\begin{document}
\title{Universal Limit on Communication}
\author{Raphael Bousso}%
 \email{bousso@lbl.gov}
\affiliation{Center for Theoretical Physics and Department of Physics\\
University of California, Berkeley, CA 94720, USA 
}%
\affiliation{Lawrence Berkeley National Laboratory, Berkeley, CA 94720, USA}
\begin{abstract}
I derive a universal upper bound on the capacity of any communication channel between two distant systems. The Holevo quantity, and hence the mutual information, is at most of order $E\Delta t/\hbar$, where $E$ is the average energy of the signal, and $\Delta t$ is the amount of time for which detectors operate. The bound does not depend on the size or mass of the emitting and receiving systems, nor on the nature of the signal. No restrictions on preparing and processing the signal are imposed. 

As an example, I consider the encoding of information in the transverse or angular position of a signal emitted and received by systems of arbitrarily large cross-section. In the limit of a large message space, quantum effects become important even if individual signals are classical, and the bound is upheld.

\end{abstract}
\maketitle

In communication theory, one typically studies problems such as signal optimization, compression, or error correction, given a particular channel. Here, I will consider a different problem: whether a channel with a desired capacity is realizable by any means, given the laws of physics. 

It will be assumed that the detection of the signal can be described using quantum field theory. A universal bound on the von Neumann entropy of quantum fields~\cite{BouCas14a} will be combined with the Holevo theorem~\cite{Hol73}. This will yield a simple, robust, and surprisingly strong bound on the information that can be conveyed between two arbitrarily large systems with arbitrary resources. The bound depends only on the energy of the signal and the length of time over which the signal can be examined.

\subsection*{Communication Between Distant Large Systems}
 
Suppose that Alice controls an arbitrarily large, bounded region of space, with arbitrary matter and energy content. For concreteness we can consider a ``planet''---an approximately spherical system of radius $R_A$---but this will not be important.  Alice would like to send a message to Bob, who resides in a distant region, outside of some much vaster sphere of radius $R_B$ (see Figures).
\begin{figure}[ht]
\includegraphics[width=0.28 \textwidth]{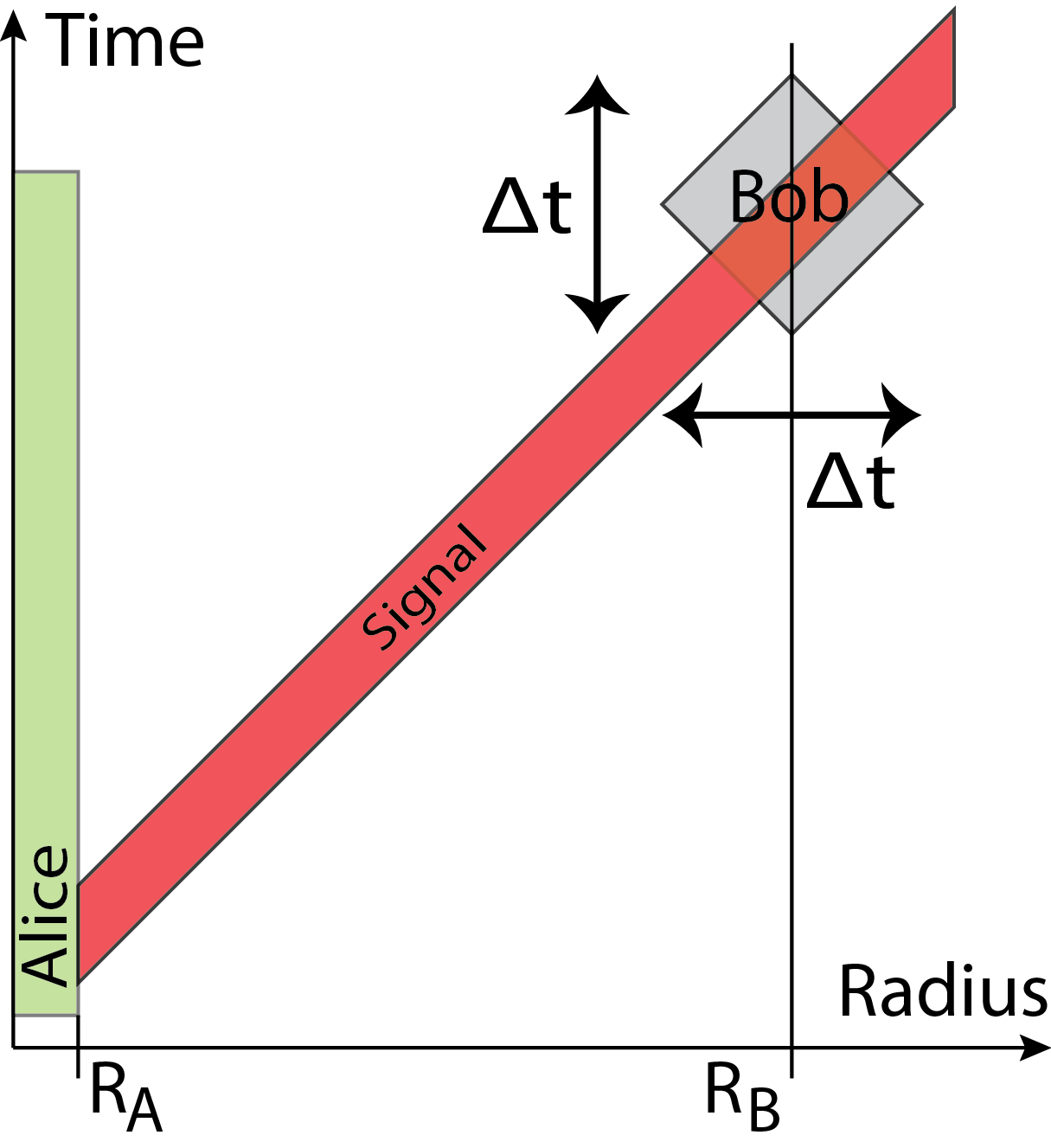}
\caption{Alice sends a signal of energy $E$ to a distant set of detectors, operated by Bob for a time $\Delta t$.}\label{fig-distant}
\end{figure}

Bob has already surrounded Alice with detectors. For example, the entire sphere at $R_B$ could be densely tiled with detectors. We require $R_B\gg R_A$ but we impose no upper limit on either $R_A$ or $R_B$. We need not assume that gravity is weak at Alice's location (though this can always be arranged by increasing $R_A$ and diluting her system). We do assume that gravity is weak at $R_B$, as would be the case for large $R_B$ in an asymptotically flat spacetime. 

Let $E$ be the energy of the signal Bob receives from Alice. ($E$ includes the rest mass, if any.) We suppose that the time period during which Bob's detectors will be operating is known to Alice, and that it has duration $\Delta t\ll R_B$. 

We do not restrict the amount of time that Alice is given to prepare her signal: she gets an arbitrarily early start. Nor do we restrict the amount of time for which Bob can process his detector output, nor the energy resources available to Alice and Bob for generating and processing the signal. With these minimal restrictions, {\em how much information can Alice send to Bob?} 

\begin{figure}[ht]
\subfigure[]{
\includegraphics[width=0.22 \textwidth]{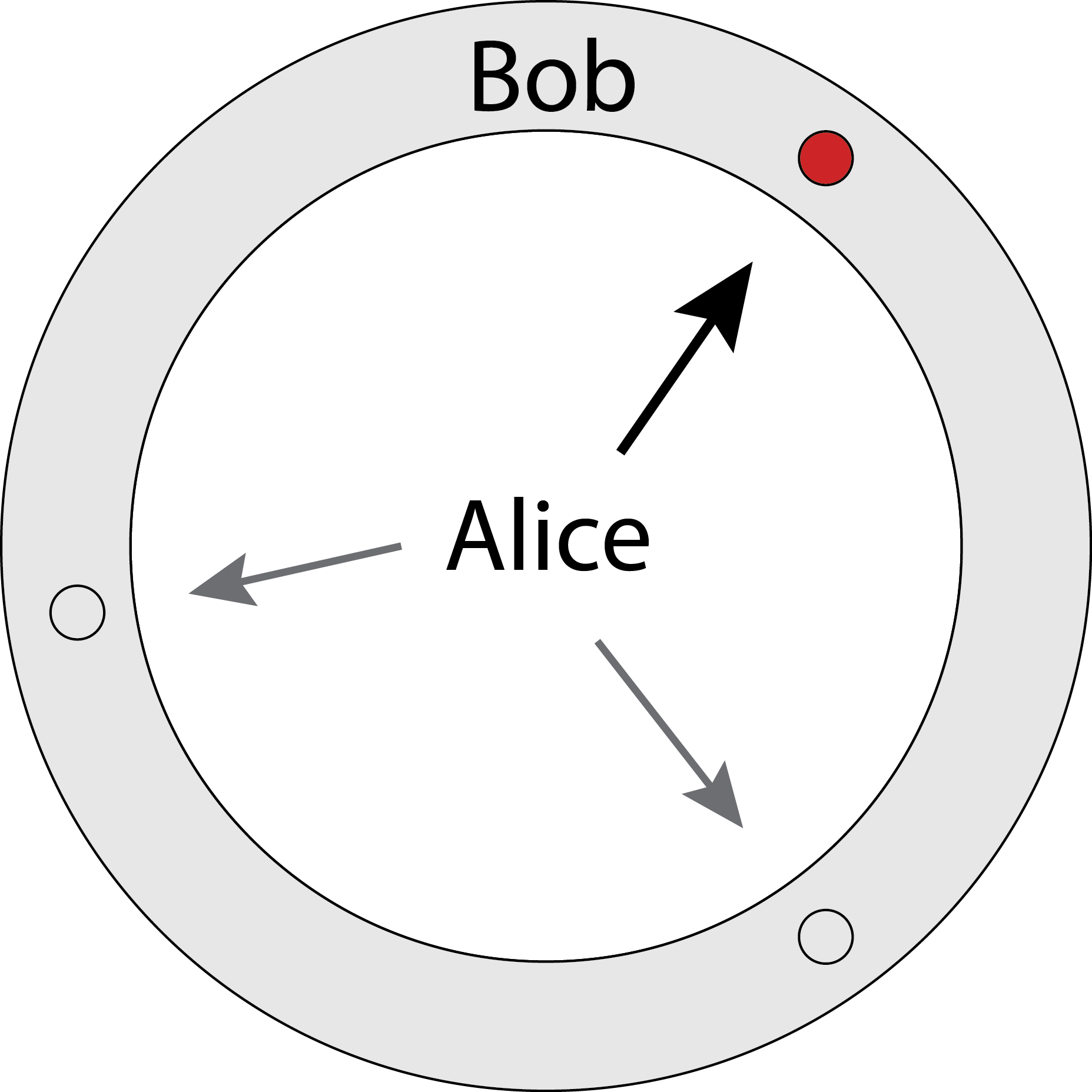}
}
\subfigure[]{
 \includegraphics[width=0.22 \textwidth]{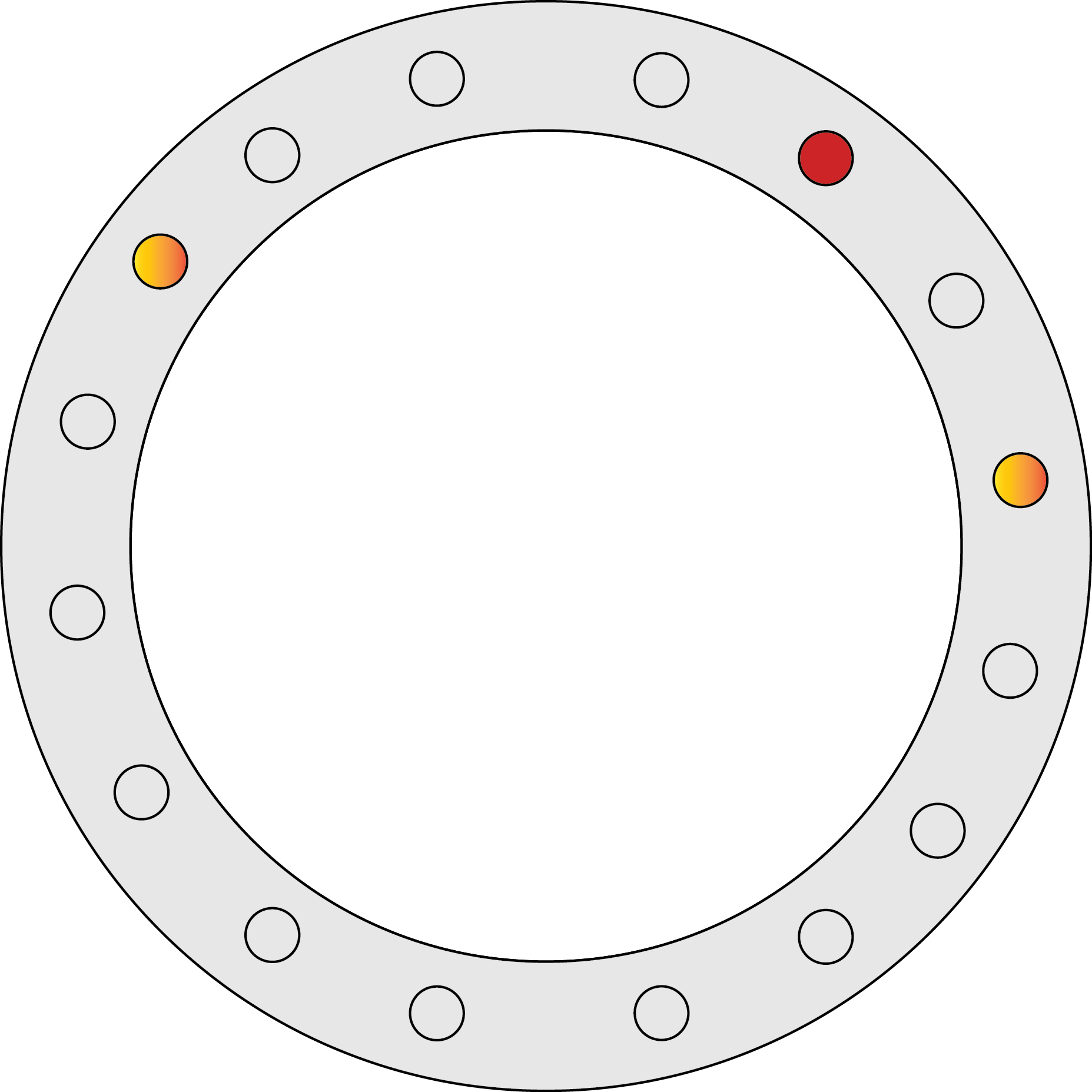}
}
\caption{(a) Small message space. Alice sends one of three previously agreed-upon classical signals (red dot) to Bob. Bob resides at a great distance $R_B$ and operates three detectors at the potential signal sites. By examining which detector responded, he learns an amount $H(A\!:\!B) = \log 3$ of information. (b) Large message space. Alice sends one of $N$ classical signals, where $N\gg E\Delta t/\hbar$. Bob's detectors can only explore a finite region (grey shell) of width $\Delta t$. Irreducible quantum noise creates false detections (yellow), which preclude Bob from identifying Alice's message.}\label{fig-bobsshell}
\end{figure}

To be precise, let us assume that they have agreed on a set of $N$ possible messages, from which Alice will select message $a$, with probability $p(a)$, to be sent to Bob as a physical signal. If Bob can distinguish reliably between all $N$ signals, and thus determine which of the $N$ possible signals actually arrived, then Bob gains an amount of information equal to the Shannon entropy of the message set, 
\begin{equation}
H(A) \equiv -\sum_a p(a) \log p(a)\leq \log N ~.
\label{eq-shannon}
\end{equation}

This can be justified as follows. Consider Bob after his detectors have received Alice's signal, but before he inspects them. At this point he may describe the message by a classical probabilistic ensemble, with Shannon entropy $H(A)$. This quantifies Bob's initial ignorance. After finding the message to be $b$, Bob updates the probability distribution to $p'(a) = \delta_{ab}$, with vanishing Shannon entropy. Thus Bob's ignorance has decreased by $H(A)$, i.e., he has gained an amount $H(A)$ of information.

With no restrictions on $N$, how much information can Bob gain, if the average signal energy is $E$, and Bob's detectors operate for a time period of duration $\Delta t$?

\subsection*{Unbounded Classical Message Space}

It would appear that Bob can gain an unbounded amount of information, because there is no limit on the number of distinct signals, at fixed average energy, such that each signal is well-localized in time to much better than $\Delta t$. 

We can construct an explicit protocol that exploits this. This is particularly simple if we use classical signals, so we shall take $E\Delta t\gg \hbar$. This ensures that Alice is able to send a classical excitation---a flash of light, say---whose duration will be short compared to $\Delta t$, or equivalently, whose spatial extent is smaller than $\Delta t$. (I use units where the speed of light is unity.) Alice sends only one such signal, encoding the message in the angular direction of the signal. Bob has detectors at all angles on his distant sphere. From the solid angle at which Bob receives the signal, he will learn what Alice's message is. 

At {\em fixed} system size $R_A$, there would be limits on the angular resolution Alice can achieve. With energy $E$, she can resolve distances $d\gtrsim \hbar/E$. The number of distinct pixels that can light up on the surface of her system is 
\begin{equation}
N\lesssim \frac{R_A^2}{d^2}\lesssim \frac{R_A^2E^2}{\hbar^2}
\label{eq-naive}
\end{equation}
Or in momentum space language, with energy $E$ Alice can excite spherical harmonics with $\ell\lesssim \ell_{\rm max}\sim  E R_A/\hbar$ on a sphere of size $R_A$. There are $\ell_{\rm max}^2$ such harmonics, leading again to Eq.~(\ref{eq-naive}). 

But the system size is not fixed. We see from Eq.~(\ref{eq-naive}) that Alice can make the message space as large as she likes, simply by emitting from a large enough sphere. It is worth stressing just how obvious this conclusion is: it follows directly from locality.  

At any fixed $N$, we may choose each message to be equally likely, $p(a) = 1/N$, so $H(A)=\log N$. Since there is no upper bound on $N$, we can make $H(A)$ as large as we like, at fixed $E,\Delta t$.

Indeed, the problem is not with Alice. Rather, it lies with Bob's ability to distinguish an arbitrary number of classical signals. This is a quantum effect, and before we can understand it, we must reformulate our classical protocol in quantum language. 

\subsection*{Quantum Description}

We will need the following standard definitions. The von Neumann entropy of a quantum state is
\begin{equation}
S(\rho) \equiv -\mathrm{tr}\, \rho\log\rho~.
\end{equation}
The relative entropy of $\rho$ with respect to $\sigma$ is
\begin{equation}
S(\rho||\sigma) \equiv \mathrm{tr}\, \rho\log\rho - \mathrm{tr}\, \rho\log\sigma~.
\end{equation}
Both are non-negative.

Alice prepares the signal state $\rho^{A}_a$ with probability $p(a)$. The state of ignorance is the average state,
\begin{equation}
\rho^{A}_{\rm av}=\sum_{a=1}^{N} p(a)\, \rho^{A}_a~,
\label{eq-globalstate}
\end{equation}
Since the signal states correspond to distinct classical states, they are mutually orthogonal as quantum states: $\rho_a\rho_{a'} = 0$ for $a\neq a'$. This implies that the von Neumann entropy of the average state is at least the Shannon entropy:
\begin{equation}
S(\rho^{A}_{\rm av})  = H(A) + \sum p(a)\, S(\rho^{A}_a) \geq H(A)~.
\label{eq-sglobal}
\end{equation}
(The inequality is saturated if all signal states are pure, $\rho^{A}_a= |a\rangle \langle a|$.) 

Bob's detectors are only operating for a time $\Delta t$. This means that he has access only to a finite region $B$: a shell of thickness $\Delta t$ bounded by two spheres of radii $R_B$, $R_B+\Delta t$. From a quantum perspective, $B$ is a subsystem (or more generally, a subalgebra) of the system $A$ which is controlled by Alice. Recall that Alice can take as much time as she likes to prepare her signal. This will be finite in practice, but since it can be much larger than $\Delta t$, we may take $A$ to be all of space (a Cauchy surface) for calculational purposes, and we shall refer to $\rho^{A}$ as a global state.

The state in the subregion $B$ accessed by Bob is fully described by the reduced density operator
\begin{equation}
\rho^{B} \equiv \mathrm{tr}_{A-B}\, \rho^{A} ~.
\label{eq-channel}
\end{equation}
The trace is over the complement of $B$, the region not probed by Bob's detectors. 

We may regard Eq.~(\ref{eq-channel}) as a quantum channel by which classical information is communicated~\cite{PreNotesCh5,MikeIke,SchWes00}. 
If Alice prepares a signal state $\rho^{A}_a$, the channel output will be $\rho^{B}_a$. Bob attempts to decode the message by performing a measurement on the system $B$. The most general measurement is described by a set of positive operators $E_i$ that sum to the identity, $\sum_i E_i = \mathbf{1}$.
The conditional probability that Bob obtains outcome $b$ is given by
\begin{equation}
p(b|a) = \mathrm{tr}_B\, (\rho^{B}_a E_b)~.
\end{equation}

\subsection*{Bounds on the Channel Capacity}

In general, $p(b|a)\neq \delta_{ab}$, which means that Bob is unable to distinguish Alice's signals perfectly. The information he gains is quantified by the classical mutual information,
\begin{equation}
H(A\!:\!B) \equiv H(A)+H(B)-H(A,B)~,
\end{equation}
which satisfies
\begin{equation}
0\leq H(A\!:\!B)\leq \mbox{min}\{H(A),H(B)\}~.
\end{equation}
Here, $H(A,B)$ is the Shannon entropy of $p(a,b)=p(a) p(b|a)$, the joint probability that Alice sends $a$ and Bob finds $b$; and $p(b) = \sum_a p(a,b)$ is the marginal (i.e., total) probability that Bob finds $b$. For example, if Bob's result is completely uncorrelated with Alice's message then $H(A\!:\!B)=0$, and Bob gains no information. If it is perfectly correlated, then $H(A\!:\!B)=H(A)$, and Bob gains all of the information about Alice's message.

For classical information sent over a quantum channel, $p(b|a)$ depends on the choice of operators $E_i$. However, $H(A\!:\!B)$ is bounded from above by the Holevo quantity~\cite{Hol73}, $\chi$, which depends only on the channel output:
\begin{equation}
\!\!\!\!\!\!\!\!H(A\!:\!B)\leq\chi \equiv S(\rho^{B}_{\rm av}) - \sum_a p(a)\, S(\rho^{B}_a) ~.
\label{eq-holevo}
\end{equation}

The von Neumann entropy of a bounded region $B$, $S(\rho^{B})$, diverges in quantum field theory, because of short-distance entanglement between excitations localized to either side of the boundaries of $B$~\cite{Sre93}. Since $\sum_a p(a)=1$, these divergences cancel in Eq.~(\ref{eq-holevo}). 

Still, it is instructive to write Eq.~(\ref{eq-holevo}) in a form where only finite quantities appear. Consider the {\em reduced vacuum},
\begin{equation}
\sigma^{B} = \mathrm{tr}_{A-B}\, \sigma^{A}~,
\label{eq-sigma}
\end{equation}
where $\sigma^{A}$ is the global vacuum state (empty Minkowski space). The {\em vacuum-subtracted entropy}~\cite{Cas08} for any quantum state in Bob's subregion is defined as
\begin{equation}
\Delta S(\rho^{B}) \equiv S(\rho^{B}) - S(\sigma^{B})~.
\end{equation}
(Vacuum subtraction is well-defined only in weakly gravitating regions such as $B$. If gravity was strong, the shape of space would depend on the quantum state. Then it would not be clear how to reduce two different states to the ``same'' region.) We now find
\begin{equation}
\!\!\!\!\!\!\!\!H(A\!:\!B)\leq\chi  =  \Delta S(\rho^{B}_{\rm av}) - \sum_a p(a)\, \Delta S(\rho^{B}_a) ~.
\label{eq-holevo2}
\end{equation}

For individual signals, which are well-localized to $B$, the vacuum-entanglement contributions cancel out, so $\Delta S(\rho^{B}_a)= S(\rho^{A}_a)$. If this remained true after averaging, we would have $\Delta S(\rho^{B}_{\rm av})= S(\rho^{A}_{\rm av})$ for the signal ensemble. We could then use Eq.~(\ref{eq-sglobal}) to recover the maximum classical channel capacity, $H(A)$, from Eq.~(\ref{eq-holevo2}).

However, I will now derive an upper bound on $\Delta S(\rho^{B}_{\rm av})$ that does not increase with the number of distinct classical signals, at fixed average signal energy. This means that for large enough $N$, $S(\rho^{B}_{\rm av})\ll S(\rho^{A}_{\rm av})$.

The log of $\sigma^{B}$ defines a {\em modular Hamiltonian} operator $\hat K$, via
\begin{equation}
\sigma^{B} = \frac{e^{-\hat K}}{\mathrm{tr}_{B}\, e^{-\hat K}}~.
\label{eq-kdef}
\end{equation} 
The {\em modular energy} of a reduced state $\rho^{B}$ is defined as
\begin{equation}
\Delta K(\rho^{B}) \equiv \mathrm{tr}_B\, \hat K \rho^{B} - \mathrm{tr}_B\, \hat K \sigma^{B}~.
\end{equation}
This quantity is useful because it allows us to trade the information theoretic quantities appearing in the Holevo bound for physical quantities. For a shell of radius $R_B$ and width $\Delta t\ll R_B$,
\begin{equation}
\Delta K(\rho^{B})\lesssim E(\rho^{B})\Delta t/\hbar~,
\end{equation}
where $E$ is the expectation value of the energy (the integrated energy density), in the state $\rho^{B}$. Precise expressions for $\Delta K$ are given in Refs.~\cite{BouCas14a,BouCas14b,Bou16}.\footnote{Refs.~\cite{BouCas14a,BouCas14b} apply to a planar array of detectors. In the limit as $R_B\to\infty$ at fixed $\Delta t$ one can also consider a spherical array~\cite{Bou16}. Other rigorous quantum entropy bounds~\cite{BouFis15a,BouFis15b,KoeLei15} may also place interesting limits on communication; I leave this question to future work.} They will not be needed here since we are interested mainly in how $H(A\!:\!B)$ scales as we increase $N$ or $H(A)$.

Positivity of the relative entropy $S(\rho^{B}||\sigma^{B})$ implies~\cite{Cas08}
\begin{equation}
\Delta S(\rho^{B}) \leq \Delta K(\rho^{B})~.
\label{eq-dsdk}
\end{equation}
Substitution into Eq.~(\ref{eq-holevo2}) yields another upper bound on the information gained by Bob,
\begin{equation}
H(A\!:\!B) \leq  \chi \leq \Delta K(\rho^{B}_{\rm av}) 
-  \sum_a p(a)\, \Delta S(\rho^{B}_a) ~. \label{eq-holevo3}
\end{equation}
This result implies that the channel capacity cannot be made arbitrarily large by enlarging the message space, at fixed average signal energy and fixed average vacuum-subtracted signal entropy $\Delta S(\rho^{A}_a)$. 

If all signals are classical and well-localized to $B$, then $\Delta S(\rho^{B}_a)\geq 0$,\footnote{If the individual signals are quantum, then {$\Delta S(\rho^{B}_a)$} can be negative. It would be interesting to constrain this regime further.} and the bound on the channel capacity simplifies to
\begin{equation}
H(A\!:\!B) \leq \chi \leq \Delta K(\rho^{B}_{\rm av}) \lesssim E_{\rm av} \Delta t/\hbar~.
\label{eq-holevo4}
\end{equation}

Eqs.~(\ref{eq-holevo3}) and (\ref{eq-holevo4}) are the main result of this paper.\footnote{Bekenstein's pioneering constraints on channel capacity~\cite{Bek81b,Bek84} used an entropy bound weaker than Eq.~(\ref{eq-dsdk}), involving the largest dimension of the problem instead of {$\Delta t$}. That upper bound would be of order {$E_{\rm av} R_B$} and thus would not constrain the channel capacity of arbitrarily large systems ({$R_B\to\infty$}). See also Ref.~\cite{AvePau14}. Bounds involving Newton's constant~\cite{Llo04} become trivial in the weakly gravitating setting considered here.}  I will turn next to its physical interpretation: for a sufficiently large message space, quantum effects become important, even if each individual signal is classical. 

\subsection*{Reduced Vacuum and Irreducible Noise}

In order to understand the bound on channel capacity at an intuitive level, let us revisit our earlier example: the signal space consists of $N$ distinct classical signals of identical energy, and $p(a) = 1/N$. Then Eq.~(\ref{eq-shannon}) implies $H(A)\to\infty$ as $N\to\infty$. But $H(A\!:\!B)$ remains bounded by Eq.~(\ref{eq-holevo4}). What prevents Bob from simply observing Alice's blatant classical signal?

To explain why Bob cannot gain unlimited amounts of information as $N\to\infty$, we adapt Casini's resolution~\cite{Cas08} of the species problem that had afflicted earlier formulations of the Bekenstein bound~\cite{Bek81}. The role of different species is played here by the different transverse or angular positions of the signal. 

By definition, Eq.~(\ref{eq-kdef}), the reduced vacuum is a thermal state with (arbitrary) temperature $\beta^{-1}$, with respect to the Hamiltonian $\beta\hat K$. Choosing $\beta = \Delta t/\hbar$, this implies that a particular signal state has nonvanishing, Boltzmann-suppressed probability of being observed in the reduced vacuum:
\begin{equation}
\log P_i \sim -\Delta K \sim -E\Delta t/\hbar~.
\end{equation}
For $E\Delta t \gg \hbar$, this probability is exponentially small for any given signal state, consistent with our intuition. 

But if Alice's sphere has many pixels, the enormous number of possible signal states can overcome the suppression, so that some false signals will appear in Bob's detectors~\cite{MarSor03,MarMin04} (Fig.~\ref{fig-bobsshell}). The expected number of false detections is
\begin{equation}
N_{\rm false} \sim N P_i~.
\end{equation}
This becomes greater than unity precisely in the regime where the bound would be violated, for $\log N\gg E\Delta t/\hbar$. In this regime, Bob will see signals that Alice did not send. Since Bob cannot determine which signal is the ``real'' one, the protocol we have devised is not obviously useful for communicating information, and so does not provide a counterexample to Eq.~(\ref{eq-holevo4}). 

(We might ask where the energy of the false flashes is coming from, $E_{\rm false}= N_{\rm false} E$. The answer is that it comes from Bob, who expends an average energy at least of order $\hbar/\Delta t$, per detector pixel, just to localize the detector operation to the time interval $\Delta t$. The total energy put in by Bob is $N\hbar/\Delta t$, which is much greater than $E_{\rm false}$ in the regime $E\Delta t \gg \hbar$.)

Alice and Bob can eliminate false signals by pruning the message space. For example, they may take only a small subset of Alice's pixels to correspond to actual messages. Then Bob does not need to operate such a large number of detectors. If $\log N\ll E\Delta t\hbar$, then it is very unlikely that even one of his $N$ detectors will produce false signals. In this regime, Eq.~(\ref{eq-holevo4}) is consistent with perfect communication, $H(A)\approx H(A\!:\!B)$.

\paragraph*{Acknowledgments} It is a pleasure to thank S.~Aaronson, H.~Casini, N.~Engelhardt, I.~Halpern, J.~Maldacena, A.~Strominger, A.~Wall, and especially Z.~Fisher for discussions.
This work was supported in part by the Berkeley Center for Theoretical Physics, by the National Science Foundation (award numbers 1521446 and 1316783), by FQXi, and by the US Department of Energy under contract DE-AC02-05CH11231.

\newpage

\bibliographystyle{utcaps}
\bibliography{all}
\end{document}